\documentclass[journal]{IEEEtran}
\IEEEoverridecommandlockouts
\usepackage{cite}
\usepackage{graphicx}
\usepackage{epstopdf}
\usepackage{amsmath,amssymb,amsfonts}
\usepackage{algorithmic}
\usepackage{textcomp}
\usepackage{xcolor}
\usepackage{pifont}
\usepackage{colortbl}
\usepackage{xcolor}
\usepackage{tikz}
\usetikzlibrary{arrows.meta, positioning, decorations.markings}
\usepackage{tabularx,booktabs}
\usepackage[flushleft]{threeparttable}

\usepackage{booktabs,ragged2e}
\usepackage{multirow}
\usepackage{caption}
\usepackage{subcaption}

\usepackage{bm}
\usepackage[colorlinks,
            linkcolor=blue,
            anchorcolor=blue,
            citecolor=blue
            ]{hyperref}
\usepackage[table]{xcolor}
\usepackage{pifont}

\newcommand{\cmark}{\ding{51}}
\newcommand{\xmark}{\ding{55}}
\newcommand{\autorefsub}[2]{%
  \hyperref[#1]{\autoref*{#1}#2}%
}

\definecolor{headergray}{RGB}{240,240,240}

\definecolor{row1}{RGB}{232,240,250} 
\definecolor{row2}{RGB}{250,238,230} 
\definecolor{row3}{RGB}{235,245,235} 
\definecolor{row4}{RGB}{242,238,248} 
\definecolor{row5}{RGB}{245,242,230} 
\definecolor{row6}{RGB}{230,245,242} 
\definecolor{rowThis}{RGB}{245,230,235} 
\definecolor{row7}{RGB}{220,235,255} 

\def\BibTeX{{\rm B\kern-.05em{\sc i\kern-.025em b}\kern-.08em
    T\kern-.1667em\lower.7ex\hbox{E}\kern-.125emX}}

\begin{document}
\bstctlcite{IEEEexample:BSTcontrol}
\title{Rethinking Secure Semantic Communications in the Age of Generative and Agentic AI: 
Threats and Opportunities}

\author{
Shunpu Tang,
Yuanyuan Jia,
Zijiu Yang,
Qianqian Yang,
Ruichen Zhang,
Jun Du, 
Jihong Park, \\
Zhiguo Shi, \IEEEmembership{Fellow, IEEE},
Jiming Chen, \IEEEmembership{Fellow, IEEE} 
\thanks{This work is partly supported by the National Key R\&D Program of China under Grant No. 2024YFE0200802, partly by NSFC under grant No.62293481, No.62201505 and No.625B2165, partly supported by the China Scholarship Council (No. 202406320381) and the CAST Young Talent Support Program for Doctoral Students. }
 \thanks{S. Tang, Y. Jia, Z. Yang, Q. Yang, Z. Shi and J. Chen are with the College of Information Science and Electronic Engineering, Zhejiang University, Hangzhou 310007, China (emails: \{tangshunpu, labulado, zijiu\_yang, qianqianyang20, shizg, cjm\}@zju.edu.cn).}
 \thanks{R. Zhang is with the College of Computing and
Data Science, Nanyang Technological University, Singapore 639798 (e-mail:
ruichen.zhang@ntu.edu.sg).}
\thanks{J. Du is with the Department of Electronic Engineering and
also with the State Key Laboratory of Space Network and Communications,
Tsinghua University, Beijing 100084, China (e-mail: jundu@tsinghua.edu.cn).}
 \thanks{J. Park is with the Information Systems Technology and Design (ISTD) Pillar, Singapore University of Technology and Design (SUTD), Singapore 487372. (Email: jihong\_park@sutd.edu.sg)}
\thanks{The corresponding author is Q. Yang.}

}

\maketitle

\thispagestyle{empty}
\pagestyle{empty}
\begin{abstract}
Semantic communication (SemCom) improves communication efficiency by transmitting task-relevant information instead of raw bits and is expected to be a key technology for 6G networks. Recent advances in generative AI (GenAI) further enhance SemCom by enabling robust semantic encoding and decoding under limited channel conditions. However, these efficiency gains also introduce new security and privacy vulnerabilities. Due to the broadcast nature of wireless channels, eavesdroppers can also use powerful GenAI-based semantic decoders to recover private information from intercepted signals. Moreover, rapid advances in agentic AI enable eavesdroppers to perform long-term and adaptive inference through the integration of memory, external knowledge, and reasoning capabilities. This allows eavesdroppers to further infer user private behavior and intent beyond the transmitted content. Motivated by these emerging challenges, this paper comprehensively rethinks the security and privacy of SemCom systems in the age of generative and agentic AI. We first present a systematic taxonomy of eavesdropping threat models in SemCom systems. Then, we provide insights into how GenAI and agentic AI can enhance eavesdropping threats. Meanwhile, we also highlight potential opportunities for leveraging GenAI and agentic AI to design privacy-preserving SemCom systems.

\end{abstract}


\section{Introduction}
Semantic communication (SemCom) has emerged as a promising paradigm for improving communication efficiency by transmitting important information relevant to receivers, rather than raw bit sequences \cite{Semantic1}. As such, SemCom is widely expected to play a key role in supporting many applications envisioned in upcoming 6G networks, such as smart cities, autonomous systems, extended reality, and the metaverse. This paradigm shift is driven by recent advances in artificial intelligence (AI), which enable SemCom to leverage end-to-end trained neural network (NN)-based semantic encoders and decoders to efficiently extract, transmit, and reconstruct semantic information. Such AI-enabled SemCom systems have demonstrated superior performance over traditional communication systems \cite{DeepJSCC,Shunpu_SemCom}. Furthermore, with the rapid development of generative AI (GenAI) techniques, SemCom systems can use powerful GenAI models trained on massive datasets to facilitate semantic encoding and decoding, thereby achieving higher communication efficiency and improved robustness under severely degraded channel conditions\cite{GenerativeAI_SemanticCom_Security_2025}.

However, such significant improvements in communication efficiency also introduce risks to information security and user privacy \cite{Secure_SemCom_IEEENetW_2024_2}. Owing to the broadcast nature of wireless channels, eavesdroppers can also benefit from advances in GenAI to reconstruct private information from corrupted and partial intercepted signals, even when their eavesdropping channels are significantly worse than those of legitimate receivers. This capability may challenge the effectiveness of classical physical-layer security (PLS) techniques, which typically rely on channel quality advantages between legitimate and eavesdropping links\cite{SemCom_Security_IEEEWC_2023}. Besides, cryptography-based approaches may not be directly applicable to SemCom systems, as their NN-based transceiver architectures are different from those used in traditional digital communication systems. Moreover, the rapidly emerging agentic AI paradigm\cite{AgenticAI_NetCom_2025} further escalates these risks, enabling eavesdroppers to automatically infer user behavior and intentions behind the transmitted data via planning, reasoning, memory accumulation, and using external knowledge. Therefore, these emerging threats motivate a rethinking of the security and privacy of SemCom systems in the age of generative and agentic AI, as well as the opportunities enabled by these technologies.
\begin{table*}[!t]
\centering
\caption{Representative tutorial and survey papers on secure SemCom in IEEE Xplore.}
\label{tab:secure_semcom_comparison}
\renewcommand{\arraystretch}{1.25}
\setlength{\tabcolsep}{5pt}
\begin{tabular}{p{3.2cm}|p{3.0cm}|p{2.1cm}|p{2.2cm}|p{2.2cm}|p{3.2cm}}
\hline
\rowcolor{headergray}
\textbf{Refs}
& \textbf{Scope}
& \textbf{Threat Modelling}
& \textbf{GenAI for Eavesdropping}
& \textbf{Agentic AI for Eavesdropping}
& \textbf{GenAI/Agentic-enabled Security Enhancement} \\
\hline

\rowcolor{row1}
IEEE WCM (2023) \cite{SemCom_Security_IEEEWC_2023}
& Novel threats in SemCom; PLS and covert communication for SemCom 
& Glass-box decoder
& \xmark
& \xmark
& \xmark \\

\hline
\rowcolor{row2}
IEEE WCM (2023) \cite{10251845}
& Secure SemCom for metaverse applications under black-box attack 
& Closed-box encoder-only
& \xmark
& \xmark
& \xmark \\

\hline
\rowcolor{row4}
IEEE Network (2024) \cite{Secure_SemCom_IEEENetW_2024}
& Multiple attack models and defense methods for SemCom
& Glass-box decoder 
& \xmark
& \xmark
& \xmark \\

\hline
\rowcolor{row3}
IEEE Network (2024) \cite{Secure_SemCom_IEEENetW_2024_2}
& Fundamentals and challenges of SemCom security
& Glass-box decoder 
& \xmark
& \xmark
& \xmark \\

\hline
\rowcolor{row1}
IEEE COMST (2025) \cite{Semantic_Security_Survey_2025}
& Architecture, security, and privacy in SemCom networks
& Glass-box decoder, Closed-box encoder-only 
& \xmark
& \xmark
& Partial (GenAI for privacy filters) \\

\hline
\rowcolor{row2}
IEEE Network (2025) \cite{GenerativeAI_SemanticCom_Security_2025}
& Opportunities and security risks of integrating GenAI with SemCom
& \xmark
&  Partial (risk discussion)
& \xmark
&  \xmark \\

\hline
\rowcolor{row4}
IEEE WCM (2025) \cite{Knowledge_Assisted_Privacy_SemCom_2025}
& Knowledge-assisted privacy preservation in SemCom
& Glass-box decoder
& \xmark
& \xmark
& Partial\\

\hline
\rowcolor{row3}
IEEE COMMAG (2025) \cite{He2024SecureSemComGenAI} 
& Improving SemCom security using GenAI models
& Glass-box decoder, Closed-box encoder-only
& \xmark
& \xmark
& Partial (GenAI for artificial noise) \\

\hline
\rowcolor{rowThis}
\textbf{This Work}
& \textbf{GenAI- and agentic-AI-enabled eavesdropping and privacy-preserving opportunities}
& \textbf{Systematic taxonomy}
& \textbf{\cmark}
& \textbf{\cmark}
& \textbf{\cmark } \\

\hline
\end{tabular}
\end{table*}

In this paper, we investigate the threats and opportunities facing SemCom systems in the age of generative and agentic AI. Compared with existing survey and tutorial papers on secure SemCom summarized in \autoref{tab:secure_semcom_comparison}, our main contributions include: (1) we present a systematic taxonomy of eavesdropping threat models in SemCom systems based on the eavesdropper's role access and model knowledge; (2) we provide insights into how generative and agentic AI enhance eavesdropping threats in SemCom systems, including private information reconstruction and inference of user behavior and intentions; (3) we discuss potential opportunities for leveraging generative and agentic AI to design privacy-preserving SemCom systems. (4) we conduct several case studies to demonstrate the emerging risks and opportunities.

\begin{figure*}[!t]
    \centering
    \includegraphics[width=1\linewidth]{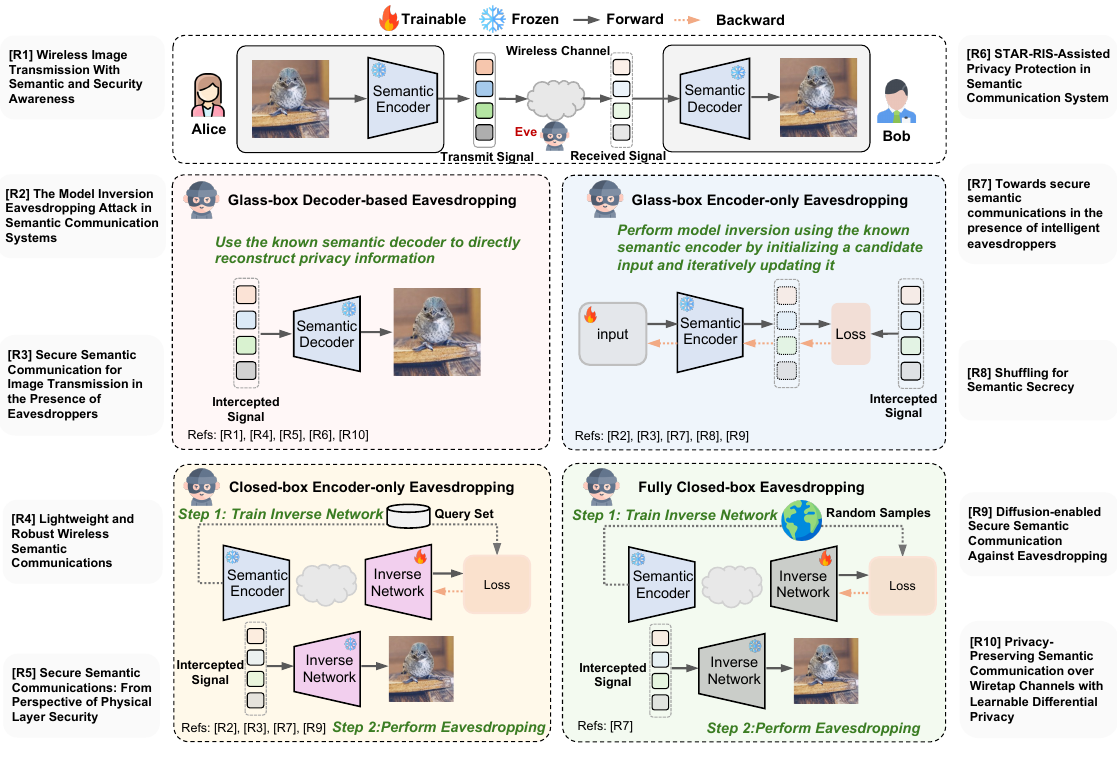}
    \caption{Illustration of eavesdropping models in SemCom systems, which summarizes four representative threat models based on the eavesdropper’s access level and model knowledge, and categorizes representative existing technical papers into each threat category.}
    \label{fig:eve_taxonomy}
\end{figure*}

\section{Eavesdropping in SemCom}
In this section, we first introduce a general framework for eavesdropping in SemCom systems and then develop a systematic taxonomy of eavesdropping threat models according to the eavesdropper’s role and model access.

\subsection{SemCom in the Presence of Eavesdroppers}
A typical SemCom system consists of a legitimate transmitter and receiver. The transmitter employs an NN-based semantic encoder that extracts high-level semantic information from the source data (e.g., image, text, speech, point cloud) and maps it into a complex-valued channel input signal. Then, the channel input is transmitted over an open wireless channel, where channel noise and fading usually exist and corrupt the signal. At the legitimate receiver, an NN-based semantic decoder is employed to reconstruct the original data from the received noisy signal with high fidelity. The semantic encoder and decoder can be jointly trained in an end-to-end manner to optimize the overall communication performance or be implemented using well-trained generative models, such as generative adversarial networks (GANs), generative diffusion models (GDMs), and large language models (LLMs).

In addition to the legitimate communication pair, eavesdroppers may also exist in this process and attempt to intercept the transmitted signal leveraging components similar to the legitimate users. This can be attributed to the several reasons: 
\begin{itemize}
\item With the progress of standardization and open-source initiatives, the architectures and parameters of semantic encoders/decoders may be publicly available.
\item In multi-user scenarios, a legitimate sender or receiver may have authorized access to the semantic models, but later misuse this access to infer private semantic information from other users' transmissions.
\item The semantic encoder and decoder may be trained with collaborative learning paradigms, such as federated learning, where malicious participants can exploit their access to the model updates.
\item Even without direct access to the semantic models, eavesdroppers can also leverage data-driven techniques to approximate the behavior of the encoder/decoder through extensive observations or queries.
\end{itemize}

\subsection{Threat Models for Eavesdropping in SemCom}
To systematically characterize the potential threats posed by eavesdroppers in SemCom systems, we classify eavesdropping threat models along the following two key dimensions:
\begin{itemize}
    \item \textbf{Role access}: This dimension describes whether the eavesdropper has previously participated in the communication system as a legitimate transmitter or receiver, thereby obtaining authorized access to the semantic encoder or decoder. 
    \item \textbf{Model access}: This describes how much the eavesdropper knows about the semantic models. Full model access means that the eavesdropper knows both the model structure and parameters. Partial access may include knowledge of the model structure or the type of training data. 
\end{itemize}
Therefore, the two dimensions can be combined to form four distinct threat models for eavesdropping in SemCom systems. We summarize these models in \autoref{fig:eve_taxonomy} and detail them in the following\footnote{In the following, we use glass-box and closed-box to refer to white-box and black-box settings, respectively.}.

\subsubsection{Glass-box Decoder-based Eavesdropping}
In this case, the eavesdropper has full access to the decoder, including its architecture and parameters, making it straightforward to reconstruct the original message by directly applying the decoder to intercepted signals. It is worth noting that, the eavesdropper typically experiences poorer channel conditions than the legitimate receiver. Nevertheless, recent advances in SemCom demonstrate that semantic decoders are robust enough and can recover important semantic information even from highly noisy or incomplete observations. This robustness exposes significant security risks when the decoder is accessible to an eavesdropper.

\begin{figure*}[!t]
    \centering
    \includegraphics[width=0.98\linewidth]{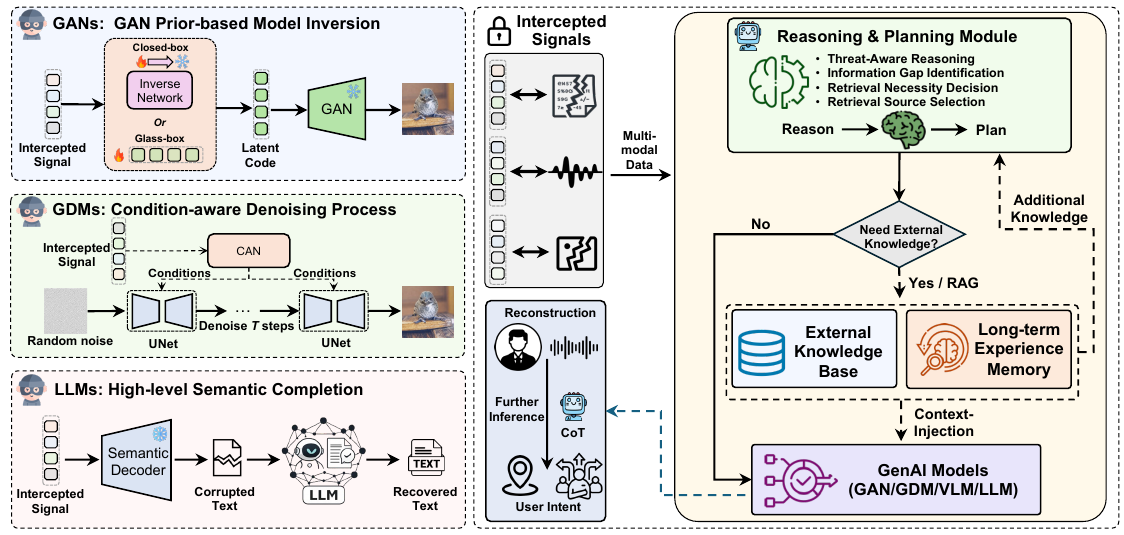}
    \caption{Illustration of GenAI-enabled eavesdropping in SemCom systems, which shows how different types of GenAI models can be exploited by eavesdroppers to enhance private information reconstruction from intercepted semantic signals. We also present an agentic AI empowered eavesdropping framework, where the eavesdropper operates as an intelligent agent that continuously observes, reasons, memorizes, and retrieves knowledge from external knowledge bases to improve private information reconstruction and further infer user behavior and intent.}
    \label{fig:eve_genAI_1}
\end{figure*}

\subsubsection{Glass-box Encoder-only Eavesdropping}
When the eavesdropper has access to the semantic encoder but not the decoder, it can still pose a serious threat to semantic privacy. In particular,  eavesdropper can  reverse the encoding process and recover sensitive semantic information from the intercepted signals. This can be typically achieved through model inversion techniques. Specifically, the eavesdropper initializes a candidate input and iteratively optimize it to minimize the difference between its derived channel input signal and the intercepted signal, thereby significantly increasing the risk of privacy leakage.

\subsubsection{Closed-box Encoder-only Eavesdropping} When the eavesdropper has no access to the internal structure or parameters of the semantic decoder, the eavesdropper must rely on indirect methods. In particular, if the eavesdropper knows the decoder type (e.g., transformer-based or convolutional NN-based), the channel characteristics, and the distribution of possible transmitted data, it can treat the decoder as a closed-box function and attempt to learn an approximate inverse mapping. This can be achieved by observing multiple transmissions over time and collecting corresponding input–output pairs under the same communication scenario. Although the learned inverse mapping may not perfectly replicate the true decoder behavior, it can still recover meaningful semantic information, posing a serious security threat, especially in long-term or large-scale communication settings where sufficient observations can be accumulated.

\subsubsection{Fully Closed-box Eavesdropping}
In this worst-case scenario, the eavesdropper has no access to the internal structures or parameters of either the semantic encoder or decoder and can only observe their input–output behavior. Moreover, the eavesdropper lacks prior knowledge of the specific model architectures and the underlying data distributions used for training, as eavesdropping may be opportunistic and leave insufficient time to acquire detailed system knowledge. Nevertheless, it remains possible to infer semantic content by blindly learning a mapping from intercepted signals to private information using data-driven techniques, alternative model architectures, and publicly available datasets.

\subsection{Lesson Learned}
Although the above threat models outline various levels of eavesdropping capabilities and techniques to intercept private semantic information, several limitations remain for practical implementation. First, due to the typical worse channel conditions experienced by eavesdroppers, the intercepted signals may be corrupted by noise and fading, making accurate reconstruction challenging. Second, the well-used model inversion might not always converge to a satisfactory solution, due to lack of sufficient observations or prior knowledge about the data distribution. Beyond these technical challenges, the current eavesdropping threat methods only try to recover the original data, while ignoring to infer the use intent behind the data. Therefore, in the next section, we will discuss how GenAI and Agentic AI can enhance the eavesdropping threats from multiple perspectives.

\section{Enhanced Eavesdropping Threats}
In this section, we first present a GenAI-empowered eavesdropping framework, illustrating how powerful generative priors enable more accurate and robust reconstruction from intercepted signals. Next, we further discuss how agentic AI enables eavesdroppers to perform long-term, adaptive inference by combining memory, external knowledge, and reasoning, allowing them to infer user behavior and intent beyond the transmitted content.

\subsection{GenAI-enabled Private Information Reconstruction}
 With the advancement of GenAI, eavesdroppers can now leverage powerful generative models as universal decoders to significantly enhance reconstruction quality, thereby posing threats to user privacy. As shown in \autoref{fig:eve_genAI_1}, we analyze the eavesdropping threats enabled by three well-known GenAI models, such as GANs, GDMs and LLMs, and discuss how they can be used to improve eavesdropping performance and therefore escalate the privacy risks in SemCom systems.
\subsubsection{GANs}
 GANs consist of a generator and a discriminator that are trained in an adversarial manner, where the generator aims to synthesize realistic data from random noise, and the discriminator tries to distinguish between real and generated data. Based on this, eavesdroppers can leverage prior knowledge learned by GANs to reconstruct high-fidelity private information\footnote{\url{https://github.com/jiupinjia/GANs-for-Inverse-Problems}}. Specifically, for the glass-box encoder scenario, the eavesdropper can optimize a latent vector input to the GAN generator such that the output matches the intercepted semantic features when passed through the known encoder. While for the closed-box encoder scenario, the eavesdropper can train an inverse mapping from intercepted signals to the GAN latent space using observed input-output pairs, and then use the trained inverse model and the GAN generator to reconstruct the original data, which makes the reconstruction more accurate and realistic.
        
\subsubsection{GDMs} GDMs generate data by reversing a gradual noise-adding and show better generation quality and diversity than GANs. As a result, eavesdroppers can leverage GDMs to further enhance eavesdropping performance. The key idea is to treat the intercepted semantic signals as noisy and compressed observations of the original data, where the degradation is introduced by the transmission process, and solve the inverse problem of this degradation. In the glass-box encoder scenario, the eavesdropper can apply the diffusion posterior sampling technique (DPS)\footnote{\url{https://github.com/DPS2022/diffusion-posterior-sampling}}, and iteratively denoise the sample, while at each step incorporating gradient information from the known encoder to ensure consistency with the intercepted signals. For the closed-box encoder scenario, the eavesdropper can also train a condition-aware network (CAN), such as ControlNet\footnote{\url{https://github.com/lllyasviel/ControlNet}}, to guide the diffusion process using the intercepted signals as external conditions.
    
\subsubsection{LLMs} LLMs pre-trained on massive text data have demonstrated remarkable capabilities in natural language understanding and generation.  For text-based SemCom systems, eavesdroppers can exploit LLMs to enhance eavesdropping performance. Specifically, in glass-box decoder scenarios, eavesdroppers can feed the decoded text containing severely corrupted or missing segments into an LLM and use carefully designed prompts to instruct the model to predict and recover the missing or corrupted content. While in closed-box encoder scenarios, techniques such as in-context learning, Adapter tuning\footnote{\url{https://github.com/AGI-Edgerunners/LLM-Adapters}}, or LoRA\footnote{\url{https://huggingface.co/docs/peft/main/en/conceptual_guides/lora}} also have potential to help eavesdroppers learn an approximate inverse mapping from intercepted signals.

\subsection{Agentic AI empowered Eavesdropping}

To further enhance eavesdropping performance, it is necessary to move beyond reconstructing private information using a single GenAI model in isolated transmissions or single-modality settings. In practical SemCom systems, intercepted transmissions are often temporally correlated, multi-modal, and task-driven. This motivates an agentic eavesdropping paradigm, in which the eavesdropper operates as an intelligent agent that continuously observes, reasons, memorizes, and adapts across multiple transmissions.

Specifically, instead of passively processing each intercepted signal independently, an agentic eavesdropper actively coordinates multiple GenAI models, internal memory, and external knowledge sources through an iterative \emph{perception--reasoning--planning--generation} loop. As illustrated in \autoref{fig:eve_genAI_1}, we present a unified framework of agentic AI based eavesdropping, which consists of the following key components and steps.

\subsubsection{Experience Memory}
In practice, semantic transmissions are not independent, as users often perform related tasks and repeatedly transmit semantically correlated content over time. Treating each intercepted signal in isolation therefore leads to suboptimal inference performance. An agentic eavesdropper maintains an experience memory that continuously records intercepted channel observations, intermediate decoding results, and reconstructed private information across multiple transmissions and modalities. 
This memory enables temporal reasoning and cumulative knowledge acquisition, allowing the eavesdropper to exploit long-term semantic correlations rather than relying on one-shot reconstruction.

\subsubsection{External Knowledge Base}
Due to limited observations and imperfect decoding, GenAI models employed by eavesdroppers, such as LLMs, may suffer from hallucination or semantic inconsistency, particularly under poor channel conditions. 
To mitigate this issue, the agent can actively query external knowledge bases, including public datasets, domain-specific repositories, and task-related knowledge graphs, to provide auxiliary semantic priors that ground the inference process. 
By incorporating externally verified knowledge, the eavesdropper improves the consistency and robustness of reconstructed private information.

\begin{figure*}[!t]
    \centering
    \includegraphics[width=0.95\linewidth]{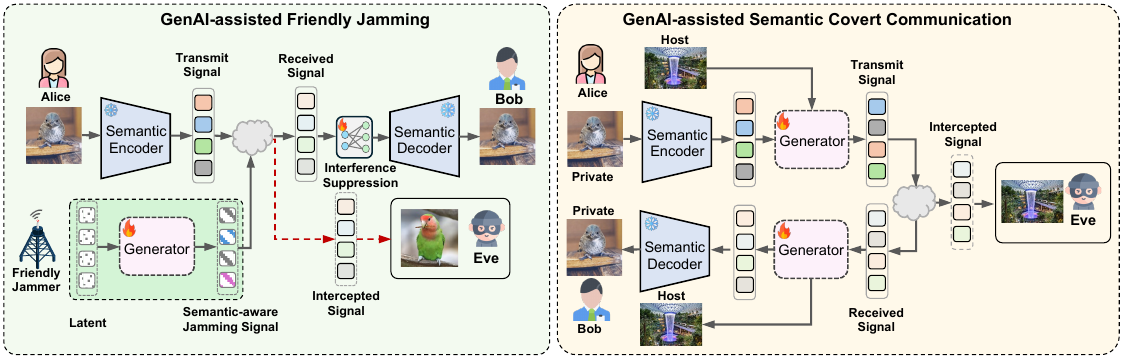}
    \caption{Illustration of typical GenAI-enabled privacy-preserving SemCom techniques, including GenAI-assisted semantic-aware friendly jamming and GenAI-assisted semantic covert communication.}
    \label{fig:genAI_secure}
\end{figure*}

\subsubsection{Reasoning and Planning Module}
A core component of agentic eavesdropping is a dedicated reasoning and planning module that orchestrates the overall inference process. 
Based on current observations and historical experience, this module performs multi-step reasoning to (i) formulate and update hypotheses regarding the transmitted content or user intent, (ii) assess uncertainty and identify information gaps, and (iii) plan subsequent inference actions, such as selecting target modalities, triggering additional knowledge retrieval, or refining generation prompts. 
This reasoning-driven planning enables adaptive decision-making across multiple transmissions, rather than single-shot naive reconstruction.

\subsubsection{Retrieval-augmented generation}
Building upon the experience memory and external knowledge base, the agent adopts RAG mechanisms, in which relevant semantic instances, contextual knowledge, or historical observations are dynamically retrieved and injected into generative models. This agent-guided generation process significantly enhances the consistency, plausibility, and reliability of reconstructed private information in complex semantic communication scenarios.

\subsubsection{Behavior and Intent Inference Module} 
Beyond reconstructing privacy-sensitive information, a critical risk posed by agentic eavesdropping is the ability to infer higher-level user behavior and intent by reasoning the intercepted signals with the help of accumulated experience memory and external knowledge. This can enable the eavesdropper to predict forthcoming user actions, such as the next task to be executed, the likely physical location, or the subsequent service or content to be requested, using techniques like chain-of-thought (CoT)\footnote{\url{https://huggingface.co/blog/samihalawa/chain-of-thoughts-guide}}, thereby facilitating proactive surveillance or further eavesdropping. Moreover, such behavior-level inference can also feedback to the reasoning and planning module, serving as high-level priors that guide the private information reconstruction process.

\begin{figure*}
    \centering
    \includegraphics[width=0.95\linewidth]{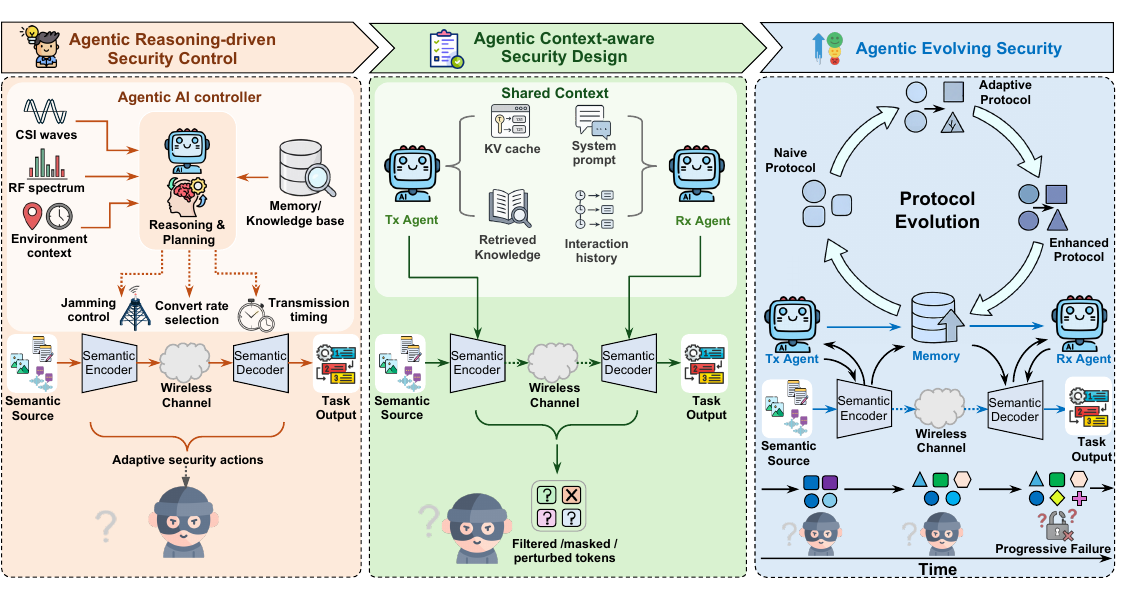}
    \caption{Perspective of agentic AI empowered privacy-preserving SemCom techniques, including agentic reasoning-driven security control, agentic context-aware security design, and agentic evolving security strategy.}
    \label{fig:agentic_secure}
\end{figure*}

\subsection{Lesson Learned}
From the above discussion, we can see that GenAI and agentic AI fundamentally enhance eavesdropping threats in SemCom systems along the following directions. 
\begin{itemize}
\item  Powerful generative priors learned from large-scale data allow eavesdroppers to significantly improve the reconstruction quality of private information, even when these signals are highly compressed, noisy, or partially missing.  As a result, traditional security assumptions that link high reconstruction distortion to low privacy risk may no longer hold.

\item By integrating experience memory, external knowledge bases, and reasoning-and-planning modules, agentic eavesdroppers are able to make a significant paradigm shift from single-shot to continuous and long-term eavesdropping, highlighting the need to consider temporal correlations and multi-modal fusion in future security strategies for SemCom.

\item Eavesdroppers no longer limited to reconstructing the original transmitted content, but further infer higher-level user behavior and intent. This significantly expands the scope of privacy risks in SemCom systems and calls for a rethinking of security metrics that consider semantic inferability, rather than only signal-level distortion.
\end{itemize}

\section{Opportunities for Privacy-Preserving SemCom}
While GenAI and agentic AI pose significant threats to the security of SemCom systems, they also offer new opportunities for designing robust defense mechanisms. Therefore, in this section, we discuss several potential and promising directions that can integrate GenAI and agentic AI to enhance privacy preservation in SemCom.

\subsection{GenAI-assisted Security Designs for SemCom}
As shown in \autoref{fig:genAI_secure}, to mitigate the enhanced eavesdropping threats posed by GenAI, a natural idea is to leverage GenAI to enhance existing security designs for SemCom, such as PLS and covert communication. In the following, we discuss these approaches in detail.

\subsubsection{Physical Layer Security}
Classical PLS techniques aim to exploit the physical characteristics of wireless channels to increase the secrecy capacity, i.e., the rate difference between the legitimate channel and the eavesdropping channel. Common methods include artificial noise generation and cooperative jamming. Therefore, GenAI can be used to generate more effective artificial noise or jamming signals that can significantly deteriorate the eavesdropping channel while minimally impacting the legitimate channel\footnote{\url{https://www.comsoc.org/publications/best-readings/physical-layer-security-foundations-intelligent-designs-implementations}}. For example, carefully designed Gaussian noise can be added to the transmitted signal, which can be viewed as a forward noising process in generative diffusion models (GDMs). By deploying GDMs at the legitimate receiver, this added noise can be effectively removed during decoding. In contrast, an eavesdropper without knowledge of the noise characteristics may fail to reconstruct the original information \cite{He2024SecureSemComGenAI}. Another approach is semantic-aware jamming, where jamming signals are designed to specifically target key semantic features (e.g., the eyes in facial images), thereby maximizing the disruption to semantic reconstruction while minimizing power consumption.

\subsubsection{Covert Communication}
Different from PLS, which focuses on degrading the decoding ability of eavesdroppers, covert communication aims to hide the existence of communication from eavesdroppers. A common approach is to jointly optimize the transmission power of the transmitter and the jammer so that the eavesdropper cannot reliably detect whether communication is occurring during a given time period. In this case, we can integrate GenAI with deep reinforcement learning (DRL) to optimize the transmission strategy \cite{Diffusion_DRl_NetOpt}, since GenAI can model complex actions and states distributions. More importantly, covert communication can be extended to the semantic level, where the goal is to conceal the presence of meaningful semantic information rather than merely the existence of signal transmission.  Specifically, a generative model with an invertible or approximately invertible structure \footnote{\url{https://github.com/XuezheMax/wolf}; \url{https://github.com/aganse/flow_models}} is used to covertly embed the channel input of a private semantic sample into that of a non-sensitive host sample. Leveraging the strong generative prior and expressive mapping capability of GenAI models, the transmitted signal retains the normal statistical and semantic characteristics of the host while implicitly carrying private information. The legitimate receiver can recover the hidden content using the shared GenAI model, whereas an eavesdropper without access to this module can only extract host-related information and remains unaware of the existence of the concealed private semantics.

\textbf{Lesson Learned:} By integrating GenAI into classical security designs, we can enhance the effectiveness of existing methods such as PLS and covert communication in SemCom systems. However, these approaches primarily focus on transceiver-level techniques, and require joint optimization of SemCom and security modules. More importantly, these techniques fail to use the contextual information in SemCom to achieve native privacy preservation. This motivates the exploration of higher-level agentic security paradigms.

\subsection{Perspective of Agentic Secure SemCom}
In addition to the above GenAI-assisted security designs for SemCom, we further envision how the rapid development of agentic AI can inspire new secure SemCom paradigms beyond transceiver-level techniques. Specifically, we discuss three promising directions as follows.

\subsubsection{Agentic Reasoning-driven Security Control}
As shown in \autoref{fig:agentic_secure}, the first direction is to incorporate agentic AI with the established security designs for SemCom systems. The key idea is to leverage the reasoning and planning capabilities of agentic AI to enable dynamic analysis and control of security risks in SemCom systems. In this direction, an eavesdropping perception agent equipped with memory and external knowledge bases can continuously monitor the physical-layer environment. By analyzing real-time channel observations, such as channel state information (CSI) and other radio-frequency features, and combining them with historical experience and external knowledge, the agent can reason the presence of active eavesdroppers in the environment. For the common passive eavesdroppers that do not emit any signals, although it is generally infeasible to directly detect their existence, the agent can also analyze the sensitivity of transmitted content, user location, communication environment to infer the potential risk of passive eavesdropping. Based on the reasoning results, the agent performs security-aware planning by deciding when, where, and how to transmit, selecting suitable security mechanisms, such as PLS-based and covert communication approaches, and adaptively optimizing their key parameters (e.g., jamming power and covert transmission rate), thereby improving overall security performance while reducing unnecessary system overhead. Compared with traditional rule-based security control methods, the agentic reasoning-driven approach can better adapt to dynamic and complex wireless environments as well as diverse communication tasks in SemCom systems.

\subsubsection{Agentic Context-aware Security Design}
Another promising direction is to leverage agentic AI to enable context-aware security designs for SemCom systems. As SemCom is increasingly moving toward token-based communication  based on large foundation models and efficient tokenization techniques to support AIGC services \cite{Token_Communications_2025}, the proper decoding and understanding of the received tokens often depend on contextual information, such as historical interactions, system prompt, retrieved knowledge, and key-value (KV)-caches\footnote{\url{https://huggingface.co/blog/not-lain/kv-caching}}. Motivated by this, we can design an agentic security framework that exploits contextual information as additional knowledge to enhance privacy preservation in SemCom systems \cite{Knowledge_Assisted_Privacy_SemCom_2025}, thereby going beyond transceiver-level techniques and eliminating the joint optimization of communication and security components. Specifically,  we can deploy two collaborative agents at the transmitter and legitimate receiver sides, respectively. The transmitter-side agent is responsible for analyzing the current communication context, and dynamically filters, masks, or perturbs sensitive semantic tokens based on the shared context with the receiver-side agent. The receiver-side agent then leverages the shared context to accurately recover the original semantic tokens from the filtered or perturbed ones.  In contrast, an eavesdropper without access to the shared context may fail to decode the transmitted tokens with proper meaning, thereby enhancing privacy preservation in SemCom systems without joint optimization of transceiver components.

\subsubsection{Agentic Evolving Security}
Based on the context-aware security design, we further envision an agentic evolving security paradigm for SemCom systems, where security mechanisms continuously evolve and adapt according to changing contexts, long-term interaction history, and accumulated agent experience. A preliminary illustration of this idea can be observed in recent agent-to-agent communication prototypes, such as Gibberlink\footnote{\url{https://github.com/PennyroyalTea/gibberlink}}, where two AI agents dynamically detect each other’s identities and switch from human natural language to a more efficient machine-oriented communication protocol that is difficult for humans to understand.  Therefore, we can further extend this idea to SemCom systems, where agents at the transmitter and receiver sides can continuously adapt and evolve their semantic encoding, transmission, and security strategies, and develop even new secure SemCom protocols that change over time, making it extremely challenging for eavesdroppers to intercept and decode the transmitted semantics. 
\begin{figure*}[!t]
    \centering
    \includegraphics[width=0.95\linewidth]{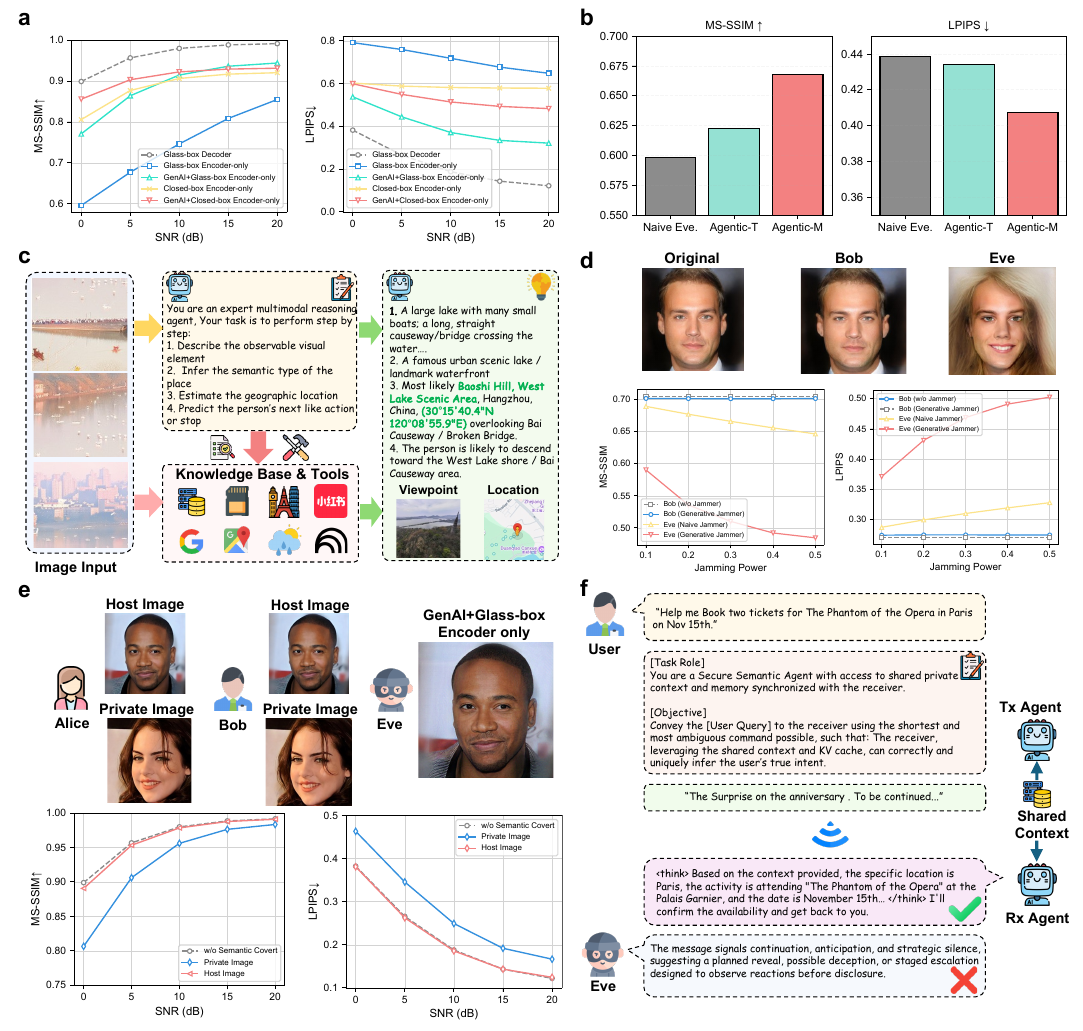}
    \caption{Case studies of GenAI- and agentic-AI-enabled eavesdropping and privacy-preserving SemCom. (a) GenAI-enabled private information reconstruction using GDMs; (b) Agentic RAG-based private information reconstruction; (c) demonstration of behavior and intent inference using agentic AI; (d) GenAI-based semantic-aware friendly jamming; (f) GenAI-assisted semantic covert communication; (e) Agentic context-aware secure SemCom framework} 
    \label{fig:case_study}
\end{figure*}

\section{Case Study}
In this section, we present several case studies to provide insights into both the eavesdropping threats and the privacy-preserving opportunities enabled by GenAI and agentic AI in SemCom systems.

\subsection{Eavesdropping Threat}

As shown in \autorefsub{fig:case_study}{a}, we first present a case study on GenAI-enabled eavesdropping threats in SemCom systems. Specifically, we consider a SemCom system for image transmission\cite{DeepJSCC} and investigate eavesdropping under the glass-box encoder and closed-box encoder-only scenarios. For GenAI model, we employ a pre-trained styleGAN2 \footnote{\url{https://github.com/NVlabs/stylegan2}} as the generative prior for eavesdropping. We compare the eavesdropping performance with and without GenAI under different channel conditions, measured by multi-scale structural similarity index (MS-SSIM) and LPIPS metrics. The results show that GenAI significantly improves the eavesdropping performance across all channel conditions. In particular, introducing GenAI into the glass-box encoder-only eavesdropping scenario yields up to 20\% gains in MS-SSIM and 0.3 reductions in LPIPS. This demonstrates the effectiveness of GenAI in enhancing eavesdropping threats in SemCom systems.

As shown in \autorefsub{fig:case_study}{b}, we present a case study on agentic AI empowered eavesdropping in SemCom systems. Following \cite{RAG_SemCom_GenAI_2025}, we consider a multi-modal SemCom setting, where the transmitter sends the semantic information of both images and image captions. The eavesdropper adopts the agentic RAG framework in \cite{RAG_SemCom_GenAI_2025} to analyze the intercepted multi-modal semantic information and retrieve relevant knowledge from an external knowledge base to reconstruct the original image.  We evaluate two agentic eavesdroppers: Agentic-T, which retrieves text-only knowledge, and Agentic-M, which retrieves multi-modal knowledge, and compare them with a non-agentic GenAI baseline. The results show that both Agentic-T and Agentic-M outperform the non-agentic GenAI eavesdropper, demonstrating the effectiveness of agentic AI in enhancing eavesdropping performance. Moreover, Agentic-M achieves the best performance, showing the benefits of multi-modal knowledge retrieval in agentic eavesdropping. 

In \autorefsub{fig:case_study}{c} we further show an example of behavior and intent inference by the agentic eavesdropper. Specifically, after reconstructing the transmitted image, the eavesdropper can further infer the user's behavior and intent using  a powerful vision-language model (VLM) named Google Gemini 3 Pro\footnote{\url{https://blog.google/technology/developers/gemini-3-pro-vision/}} with CoT prompting. As well, kinds of knowledge bases and tools, such as landmark database and Google Maps can be integrated to support the inference process. The results show that, even from  partially reconstructed image, the eavesdropper can successfully infer the geographic location, retrieve the corresponding street view, and further predict the user’s potential next actions by jointly reasoning over contextual cues, nearby points of interest, and typical behavior patterns. This highlights the significant privacy risks posed by agentic eavesdropping in SemCom systems.

\subsection{Opportunities for Privacy-Preserving SemCom}

As shown in \autorefsub{fig:case_study}{d}, we present a case study on GenAI-assisted PLS for SemCom systems. We consider a GenAI-assisted friendly jamming scheme, where a GAN is used to generate semantic-aware jamming signals, and a refined module is employed at the legitimate receiver to mitigate the impact of jamming. Without loss of generality, the transmit power of the legitimate transmitter is normalized to 1, where we vary the jamming power from 0 to 0.4. Representative reconstructed images are shown at the top, while quantitative results in terms of MS-SSIM and LPIPS are reported at the bottom. The results indicate that, as the jamming power increases, the reconstruction quality at the legitimate receiver remains nearly unchanged, demonstrating the effectiveness of the refinement module. In contrast, Eve experiences a significant degradation in semantic reconstruction quality, with MS-SSIM decreasing and LPIPS increasing monotonically. Moreover, compared with naive jamming that generates random Gaussian noise, the proposed GenAI-assisted jamming scheme achieves much better secrecy performance under the same jamming power. This highlights the potential of GenAI in enhancing PLS for SemCom systems.

\autorefsub{fig:case_study}{e} demonstrates a case study on semantic covert communication, where two identical invertible GenAI models are employed at the transmitter and legitimate receiver for performing signal steganography and de-steganography, respectively, and channel SNR varies from 0 dB to 20 dB. From the quantitative results, we can see that the legitimate receiver can achieve comparable reconstruction performance as the baseline SemCom system without any security design. In contrast, from the visual results at the top, we can see that Eve can only recover the host-related information and remains unaware of the existence of the concealed private semantics. This further demonstrates the great potential and opportunities of GenAI in enhancing privacy preservation for SemCom systems.

\autorefsub{fig:case_study}{f} presents a case study on agentic context-aware secure SemCom framework. Specifically, we consider the scenario where user equipped with Tx Agent wants to send sensitive instructions to Rx Agent at receiver side while preventing Eve from eavesdropping.  The Tx Agent will analyze the current communication context, and rewrite the sensitive instructions into a context-free version before transmission. The Rx Agent then leverages the shared context to accurately recover the original instructions from the rewritten version. From the results, even though the eavesdropper intercepts the rewritten instructions, it fails to infer the true meaning without access to the shared context, thereby can not perform next-step actions. In contrast, the Rx Agent can successfully understand the booking instructions and execute the task accordingly after thinking with the help of shared context. Although this example serves as a preliminary demonstration, it highlights the strong potential of agentic AI for enabling context-aware and privacy-preserving SemCom systems.


\section{Conclusion}
In this paper, we have discussed the significant threats posed by GenAI and agentic AI to the security of SemCom systems as well as the new opportunities they offer for privacy-preserving SemCom. Specifically, we first gave an overview of eavesdropping threat models. Then, we presented how GenAI enhances eavesdropping threats. We further presented an agentic AI empowered eavesdropping framework, where the eavesdropper operates as an intelligent agent to improve private information reconstruction and further infer user behavior and intent. In addition, we explored several potential directions for integrating GenAI and agentic AI to enhance privacy preservation in SemCom systems, including GenAI-assisted PLS and covert communication, as well as agentic-AI based secure SemCom framework. Finally, we presented case studies to validate our insights and illustrate both the emerging threats and the potential defenses approaches.
\bibliographystyle{IEEEtran}
\bibliography{references}
\end{document}